\documentstyle[12pt]{article}
\newcommand{\A}{{\cal A}}
\newcommand{\bmu}{{\wb \mu}}
\newcommand{\wb}{\bar}

\newcommand{\be}{\begin{equation}}
\newcommand{\ee}{\end{equation}}
\newcommand{\ben}{\begin{eqnarray}\displaystyle}
\newcommand{\een}{\end{eqnarray}}
\newcommand{\refb}[1]{(\ref{#1})}
\newcommand{\p}{\partial}

\begin{document}

{}~ \hfill\vbox{\hbox{hep-th/9510229}\hbox{MRI-PHY/23/95}}\break

\vskip 3.5cm

\centerline{\large \bf A NOTE ON MARGINALLY STABLE BOUND STATES}
\centerline{\large\bf IN TYPE II STRING THEORY}

\vspace*{6.0ex}

\centerline{\large \rm Ashoke Sen\footnote{On leave of absence from Tata
Institute of Fundamental Research, Homi Bhabha Road, Bombay 400005, INDIA}
\footnote{E-mail: sen@mri.ernet.in, sen@theory.tifr.res.in}}

\vspace*{1.5ex}

\centerline{\large \it Mehta Research Institute of Mathematics}
 \centerline{\large \it and Mathematical Physics}

\centerline{\large \it 10 Kasturba Gandhi Marg, Allahabad 211002, INDIA}

\vspace*{4.5ex}

\centerline {\bf Abstract}

Spectrum of elementary string states in type II string theory contains
ultra-short multiplets that are marginally stable. $U$-duality
transformation
converts these states into bound states at threshold of $p$-branes
carrying Ramond-Ramond charges, and wrapped around $p$-cycles of a torus.
We propose a test for the existence of these marginally stable bound
states. Using the recent results of Polchinski and of Witten,
we argue that the spectrum of bound states of $p$-branes is in agreement
with the prediction of $U$-duality.

\vfill \eject

The spectrum of type IIA or IIB string theory compactified on a torus
contains single particle
states in the ultra-short (256 dimensional) multiplet.
These states
are characterized by the following two properties: (1) the left and
the right moving charges carried by these states are equal in magnitude,
and (2) both,  the left and the right moving oscillators are in their
ground
state. All such states come
with unit multiplicity, {\it i.e.} for a given set of charge quantum
numbers satisfying the property (1) above, there is only one
ultra-short multiplet. Some of these states are
marginally stable, {\it i.e.} it is energetically possible for these
states to decay into two or more single particle states at rest\cite{DH}.
This
happens if the electric charge vector, represented by an element of the
charge lattice, is an integral multiple of another vector in the lattice.
Examples of such states are 1) the ground state of a string wrapped $m$
times around one of the cycles on the torus, 2) the Kaluza-Klein modes of
the massless states of the ten dimensional theory carrying $m$ units of
momentum along one of the compact directions, etc.

Consider now the type II theory compactified on a $p$ dimensional torus
$T^p$.
We shall choose $p$ to be odd for the type IIB theory and even for the
type IIA theory, for reasons that will become clear soon. In this case we
can consider a supersymmetric soliton solution of the theory that
corresponds to a
$p$-brane carrying Ramond-Ramond (RR) charge\cite{HOST} (which exists in
the IIA theory for even $p$ and in the IIB
theory for odd $p$) wrapped $m$-times around the $p$-cycle of the torus.
For
definiteness we shall choose $x^\mu$ for $(10-p)\le \mu \le 9$ to be the
compact directions. If we denote by $\A^{(p+1)}_{\mu_1\cdots \mu_{p+1}}$
the $(p+1)$-form gauge field arising in the RR sector,
then the $p$-brane described above is charged under the $(10-p)$
dimensional gauge field $\A^{(p+1)}_{(10-p)\cdots 89 \bmu}$ where $x^\bmu$
denotes one of the non-compact directions ($0\le \bmu\le (9-p)$).
In fact, this solution carries $m$ units of $\A^{(p+1)}_{(10-p)\cdots 89
\bmu}$ charge. We shall denote by $Q^{(p+1)}$ the charge associated with
the field $\A^{(p+1)}_{(10-p)\cdots 89 \bmu}$, normalized so that
$Q^{(p+1)}$ is always an integer. The question that we shall
be interested in is: `Are there
any ultra-short multiplets in the sector $Q^{(p+1)}=m$; and if so,
how many of them are present for a given value of $m$?'

A prediction for the degeneracy of these states is given
by the
general $U$-duality conjecture of Hull and Townsend\cite{HULL}. It can be
easily seen that by combining the $T$-duality transformations (which
sometimes convert a type IIB theory to a type IIA theory and vice versa)
with the SL(2,Z) symmetry of the ten dimensional type IIB theory, we can
transform the field $\A^{(p+1)}_{(10-p)\cdots 89 \bmu}$ to $B_{9\bmu}$,
where $B_{\mu\nu}$ denotes the rank two anti-symmetric tensor field arising
in the NS-NS sector.\footnote{Consider, for example the duality
transformation that inverts the radii of the $(p-1)$ circles labelled by
coordinates
$x^{10-p}, \ldots x^8$. This would convert $\A^{(p+1)}_{(10-p)\cdots 89
\bmu}$ to $B'_{9\bmu}$ where $B'_{\mu\nu}\equiv \A^{(2)}_{\mu\nu}$ is the
rank two anti-symmetric
tensor field in the IIB theory arising in the RR sector. With the help of
the conjectured SL(2,Z) symmetry transformation in the ten dimensional
type IIB string theory, we can convert $B'_{9\bmu}$ to $B_{9\bmu}$.}
Since $B_{9\bmu}$ couples to the winding number of the fundamental string
in the $9th$ direction, we see that the duality transformation discussed
above transforms a state of $Q^{(p+1)}=m$ to a state of the fundamental
string winding $m$ times around the $9th$ direction. Since we already know
that a fundamental string winding $m$ times around $x^9$ has in its
spectrum an ultra-short multiplet, there must also be an ultra-short
multiplet with $Q^{(p+1)}=m$ if $U$-duality is a valid symmetry of string
theory. In future we shall refer to ultra-short multiplets as BPS states.

The BPS states with $Q^{(p+1)}=m$ can be thought of as representing bound
states of $m$ elementary $p$-branes, each
wrapped only once around the $p$-cycle. Studying the existence of these
bound states is made difficult by the fact that they have zero binding
energy. In other words there is no energy barrier against pulling the $m$
$p$-branes away from each other. A related problem is that since these
states have exactly the same energy as $m$  BPS states $-$ each with
$Q^{(p+1)}=1$ $-$  at rest, it is difficult to separate this single
particle state from the continuum.\footnote{ However, in $N=2$
supersymmetric gauge
theories this difficulty has been circumvented, and one does find
such marginally stable bound states, as
predicted by duality\cite{N=2}.} For these reasons we shall adopt an
indirect approach for studying the existence of these bound states.

Suppose an ultra-short multiplet carrying $Q^{(p+1)}=m$ does exist in the
$(10-p)$ dimensional theory obtained by compactifying the type II theory
on $T^p$. Let us now compactify one more direction (which we shall take
to be the direction $x^1$ for definiteness) on a circle of radius $R$.
Then the original ultra-short multiplet will give rise to an infinite
number of ultra-short multiplets in this $(9-p)$ dimensional theory, $-$
one for each value of the internal momentum along the $x^1$ direction. Let
us denote by $P^1$ the internal momentum along $x^1$, normalized so that
it is always an integer, and focus on a state with $P^1=n$, where $n$ is any
arbitrary integer. If we choose $n$ in such a way that $n$ and $m$ are
relatively prime, then the ultra-short multiplet in the
$(Q^{(p+1)}=m,P^1=n)$ sector
becomes {\it absolutely stable},
even though the original ultra-short multiplet
in the $(10-p)$ dimensional theory was only marginally stable. Thus one of
the consequences of the existence of marginally stable ultra-short
multiplets in $(10-p)$
dimensional theory, carrying $Q^{(p+1)}$ charge $m$, is that {\it in the
$(9-p)$
dimensional
theory there must exist absolutely stable ultra-short multiplets with
$Q^{(p+1)}=m$, and $P^1=n$, for every
integer $n$ for which
$(m,n)$ are relatively prime.} The existence of the latter states
will be much easier to verify. In fact, these states can be
mapped to the winding modes of the $(m,n)$ string discussed in
refs.\cite{SCHWARZ,WITTENNEW} via $T$-duality transformation. Consider for
example the $T$-duality transformation that inverts the radii of all the
$p+1$ circles labelled by $x^1, x^{(10-p)}, \ldots x^9$. This transforms
$G_{1\mu}$ to $B_{1\mu}$ and $\A^{(p+1)}_{(10-p)\ldots 9\bmu}$ to
$B'_{1\bmu}\equiv \A^{(2)}_{1\bmu}$. Thus the $(Q^{(p+1)}=m, P^1=n)$ state
is mapped to an $(m,n)$ string wrapped around the $x^1$ direction. The
existence of $(m,n)$ strings for relatively prime $m,n$ has already been
shown in ref.\cite{WITTENNEW}; this then guarantees the existence of
$(Q^{(p+1)}=m, P^1=n)$ states. We shall now see explicitly how it works.

It has recently been shown by Polchinski\cite{POLNEW}
that there is an exact conformal field theory description of a $p$-brane
carrying RR charges. This is given by a Dirichlet brane, $-$ a $p$
dimensional hypersurface such that open strings satisfy Dirichlet boundary
conditions for coordinates transverse to the hypersurface, and Neumann
boundary condition for coordinates tangential to the hypersurface. Let us
apply this to the case of a static $p$-brane wrapped once around a $p$-torus
labelled by the coordinates $x^\mu$ for $(10-p)\le \mu\le 9$, situated at
the point $x^\mu=a^\mu$ for $1\le\mu\le (9-p)$. The open strings then satisfy
the boundary condition:
\ben \label{e1}
\p_\sigma X^\mu & = & 0 \qquad \hbox{for} \, \, (10-p)\le \mu \le 9, \cr
X^\mu & = & a^\mu \qquad \hbox{for} \, \, 1\le \mu \le (9-p), \cr
\p_\sigma X^0 & = & 0\, ,
\een
where $\sigma$ denotes the coordinate along the length of the string. If
we further compactify the coordinate $x^1$ along a circle of radius $R$,
then the open string boundary condition on $X^1$ will be modified to,
\be \label{e2}
X^1 = a^1 \qquad \hbox{mod} \qquad 2\pi R\, .
\ee
The generalization to the case of multiple $p$-branes is straightforward.
If $\{a_i^\bmu\}$ ($1\le i\le m$, $1\le \bmu\le (9-p)$) denote the
coordinates of the $m$
$p$-branes, then we have $m^2$ different sectors of open string states,
with each end being allowed to lie on any of the $m$ $p$-branes.

We want to study if the quantization of the classical solution
representing this multi- $p$-brane solution gives rise to a BPS state
carrying $n$ units of momentum in the $x^1$ direction. For this it will be
convenient to map this problem to another problem by a $T$-duality
transformation that inverts the radius $R$ in the $x^1$ direction. This
has several effects: 1) it converts a type IIB theory to a type IIA theory
and vice versa, 2) it converts the Dirichlet boundary condition on the
coordinate $X^1$ to Neumann boundary condition\cite{POLOLD}, 3) it
converts the
quantum number representing momentum along $x^1$ to winding number, and 4)
it converts $\A^{(p+1)}_{(10-p)\ldots 9\bmu}$ to
$\A^{(p+2)}_{1(10-p)\ldots 9\bmu}$, and hence $Q^{(p+1)}$ to $Q^{(p+2)}$.
Thus if we denote by $y^\mu$ the space-time coordinates in the transformed
theory, then the new conformal field theory is
described by the following open string boundary conditions:
\ben \label{e10}
\p_\sigma Y^\mu & = & 0 \qquad \hbox{for} \, \, (10-p)\le \mu \le 9, \cr
\p_\sigma Y^0 & = & 0\, , \cr
\p_\sigma Y^1 & = & 0\, , \cr
Y^\mu & = & a_i^\mu \qquad \hbox{for} \, \, 2\le \mu \le (9-p),
\een
where $1\le i\le m$. $y^1$ is a compact coordinate with radius $R^{-1}$,
whereas $y^\mu$ for $(10-p)\le \mu\le 9$ are compact coordinates labelling
the original torus $T^p$. The rest of the $y^\mu$ are
non-compact.\footnote{Note that the parameters $\{a_i^1\}$ have
disappeared
from eq.\refb{e10}, but they will reappear as components of the gauge field
along the $y^1$ direction when we take into account the collective
coordinates of this solution.}
This conformal field theory represents $m$ $(p+1)$-branes, each wrapped
around the $p+1$ torus $T^{p+1}$
labelled by the coordinates $y^1$, and $y^\mu$ for $(10-p)\le \mu \le 9$.
We want to look for BPS states of this system which carry $n$ units of
winding number along the $y^1$ direction. For this we need to know
the dynamics of collective coordinates of this system. This has been given
recently in a beautiful paper by Witten\cite{WITTENNEW} where he shows
that the low energy dynamics of this system is described by a
supersymmetric $U(m)$ gauge theory in $(p+2)$ dimensions, obtained by
dimensional reduction of the $N=1$ supersymmetric $U(m)$ gauge theory in
10 dimensions. The base manifold here is $R\times T^{p+1}$ where $R$ is
labelled by the time coordinate $y^0$, and $T^{p+1}$ by the coordinates
$y^1$ and $y^\mu$ ($(10-p)\le \mu \le 9$). Furthermore a state carrying
$n$ units of winding along the $y^1$ direction corresponds to a state
in this $U(m)$ gauge theory characterized by $n$ units of $U(1)$ electric
flux along the $y^1$ direction. (This in turn implies that there is also
$SU(m)$ electric flux along the $y^1$ direction in the representation
corresponding to the anti-symmetric product of $n$ fundamental
representations.)

Since we are interested in only the BPS states of the system, and
in only those states that do not carry any quantum
numbers representing
momentum and winding in the directions $y^\mu$ for $(10-p)\le \mu\le 9$,
we can
ignore fluctuations in the fields that depend non-trivially on any of these
directions. In other words, for our purpose {\it we can consider dimensional
reduction of the $(p+2)$ dimensional theory to a two dimensional theory
labelled by the coordinates $y^0$ and $y^1$.} This corresponds to $N=8$
supersymmetric $U(m)$ gauge theory in 1+1 dimensions, and is precisely the
theory that has been analyzed in ref.\cite{WITTENNEW}.\footnote{Although
it seems that the theory we have obtained this way is independent of $p$,
there is, in fact, a subtle dependence on $p$. The theory has eight matrix
valued scalar fields in the adjoint representation of $U(m)$. $p$ of these
fields come from the internal components of the gauge fields in the $p+2$
dimensional theory. It was shown in ref.\cite{WITTENOLD} that some of the
directions in the field space (corresponding to the internal components of
the gauge fields) need to be periodically identified.
This introduces a dependence of the theory on $p$.
Since we are interested only in the question of existence of
supersymmetric ground states in the theory, and since the supersymmetric
ground states of the theory found in ref.\cite{WITTENNEW} are well
localized in the field space, we expect that the extra periodic
identification in the field space will not destroy the
supersymmetric ground states found in \cite{WITTENNEW}.}
Following
\cite{WITTENNEW} we see that the problem of proving the existence of the
required bound state reduces to the problem of proving the existence of a
supersymmetric ground state in the N=8 supersymmetric $SU(m)$ gauge theory
in two dimensions, in the sector that carries an electric flux along the
space direction $y^1$ (which in this case is compact) in the
representation corresponding to
anti-symmetric product of $n$ fundamental representations of $SU(m)$. It
has already been argued
in ref.\cite{WITTENNEW} that such states do exist for every pair of
integers $(m,n)$ which are relatively prime.
This in turn establishes the
existence of the $p$-brane bound states in the $(9-p)$ dimensional string
theory, carrying $m$ units of $Q^{(p+1)}$
charge, and $n$ units of momentum along the $x^1$ direction, for every
pair of relatively prime integers $(m,n)$.

As has been argued before, this
is a necessary condition for the existence of a marginally stable
$p$-brane bound state in $(10-p)$ dimensions, carrying  $m$ units of
$Q^{(p+1)}$ charge.
One might ask if it is possible to see
these bound states by working directly in the $(10-p)$ dimensional theory.
In
this case, there is no need to compactify the $x^1$ coordinate, and we can
work directly with the Dirichlet $p$-brane described in eq.\refb{e1}. The
low energy dynamics of $m$ such $p$-branes is given, according to
ref.\cite{WITTENNEW}, by a supersymmetric $U(m)$ gauge theory in $p+1$
dimensions, obtained by dimensional reduction of the $N=1$ supersymmetric
$U(m)$ gauge theory in 9+1 dimensions. The base space of this $p+1$
dimensional
theory is $T^p\times R$, labelled by the space coordinates $x^\mu$ for
$(10-p)\le \mu\le 9$, and the time coordinate $x^0$. As in
ref.\cite{WITTENNEW}, the $U(1)$ part of the $U(m)$ theory is responsible
for describing the overall center of mass motion of the state, as well as
for producing the 256 fold degeneracy that is appropriate for an
ultra-short multiplet. Thus the non-trivial information comes from the
$SU(m)$ part of the theory, and the problem of counting the number of
ultra-short multiplets reduces to the problem of counting the number of
supersymmetric ground states of the supersymmetric $SU(m)$ gauge theory on
$T^p\times R$.

In this case since we are interested in states which carry only the charge
$Q^{(p+1)}$, and no other winding or momenta in any of the internal
directions, the sector of the supersymmetric gauge theory that we need to
analyze does not carry any background $SU(m)$ electric field. As a result,
for studying supersymmetric ground states of the theory we can ignore
fluctuations in the fields that depend on any of the $p$ compact
coordinates and consider dimensional reduction of the theory to
(0+1) dimensions. Thus the problem under consideration reduces to a
supersymmetric
quantum mechanics problem obtained by dimensionally reducing the $N=1$
supersymmetric Yang-Mills theory from (9+1) dimensions to (0+1) dimension.
The number of normalizable supersymmetric ground states of this system
will be in one to one correspondence with the number of ultra-short
multiplets in the type II theory on $T^p$ carrying $Q^{(p+1)}$ charge $m$.
Since $U$-duality predicts that the latter number is one, we arrive at the
following conclusion:
\noindent {\it  The
quantum mechanical system, obtained by dimensional reduction of the $N=1$
super Yang-Mills theory with gauge group $SU(m)$ from (9+1) dimension to
(0+1) dimension, should
have a unique normalizable supersymmetric ground state.}

Explicit verification of this prediction is made complicated by the
fact that there is no energy barrier that prevents a zero energy state
from spreading
out to infinity. As a result, even if a normalizable super-symmetric
ground state exists, it will only
decay according to a power law, and not exponentially for large
separation. Note again that a subtle dependence on $p$ arises from the
fact that for non-zero $p$ some of the directions in the space of scalar
fields need to be periodically identified.

For $p=0$, the marginally stable states that we are looking for correspond
to point like states in 10
dimensional type IIA string theory, carrying $m$ units of $A_\mu\equiv
\A_\mu^{(1)}$  charge\cite{WITTENNEW}. The existence of these states is
essential for
establishing that the strong coupling limit of the type IIA theory in ten
dimensions is described by the 11-dimensional supergravity theory
compactified on a circle of large radius\cite{WITTEN}.

It should be possible to carry out a similar analysis for more general
compactification of the type II theory, on $K3$ surfaces or on Calabi-Yau
3-folds. The expected degeneracy of ultra-short multiplets for
compactification of the type IIA / IIB theory on $K3$ surfaces may be
worked out in some cases
using the string-string duality conjecture\cite{HULL} / the SL(2,Z)
symmetry of the ten dimensional type IIB theory. For example, if we
consider a two-brane wrapped $m$ times around a two cycle of $K3$ that
does not self
intersect, then the BPS states in this sector will be mapped to the BPS
states in the heterotic string theory with $\vec Q_L^2=\vec Q_R^2$
($N_L=1$). In this case we expect 24 (16-dimensional) BPS
super-multiplets for every integer $m$. For two branes wrapped around two
cycles which self-intersect, the answer will be more complicated. For
compactification
on Calabi-Yau 3-folds, one can get information about the expected
degeneracy of the short multiplets by  either requiring consistent
resolution of the conifold singularity\cite{STROM} or by using
string-string duality conjecture\cite{KACHRU}. It will be interesting to
see if one can explicitly verify some of these predictions.

I wish to thank J. Schwarz for useful correspondence.

\end{document}